\input{aipcheck}

\documentclass[
    ,final            
  ]
  {aipproc}

\usepackage{url}
\usepackage{graphicx}
\usepackage{amsmath}
\usepackage{amssymb}

\usepackage{makeidx}

\layoutstyle{6x9}

\newcommand{\be}{\begin{equation}}
\newcommand{\ee}{\end{equation}}
\newcommand{\bea}{\begin{eqnarray}}
\newcommand{\eea}{\end{eqnarray}}
\newcommand{\bi}{\begin{itemize}}
\newcommand{\ei}{\end{itemize}}
\newcommand{\ben}{\begin{enumerate}}
\newcommand{\een}{\end{enumerate}}

\newcommand{\lc}{\left[}
\newcommand{\rc}{\right]}

\def\frac#1#2{{{#1}\over {#2}}}
\def\gsim{\mathrel{\rlap{\lower4pt\hbox{\hskip1pt$\sim$}}
    \raise1pt\hbox{$>$}}}         
\def\lsim{\mathrel{\rlap{\lower4pt\hbox{\hskip1pt$\sim$}}
    \raise1pt\hbox{$<$}}}         

\newcommand{\draft}[1]{}

\begin{document}

\title{Nuclear parton distributions and deviations from 
DGLAP at an Electron Ion Collider}


\author{Alberto Accardi}{
  address={Hampton University, Hampton, VA 23668, USA},
  altaddress={Jefferson Lab, Newport News, VA 23606, USA}
}

\author{Vadim Guzey}{
  address={Jefferson Lab, Newport News, VA 23606, USA}
}

\author{Juan Rojo}{
  address={Dipartimento di Fisica, Universit\`a di Milano and
INFN, Sezione di Milano, Italy}
}

\begin{abstract}
We explore 
the potential of 
an 
Electron Ion Collider to determine nuclear
modifications of PDFs. 
We find that gluon shadowing can be accurately measured down
to $x=10^{-3}$,
and discuss the possibility of detecting  non--linear QCD
effects with inclusive measurements.
\end{abstract}

\maketitle


One of
the main physics goals of 
a  future
 Electron Ion Collider (EIC) will be to accurately measure
nuclear modifications of gluons and quarks as well as
the possible onset of non-linear QCD dynamics
in heavy nuclei.
In this contribution we present a preliminary
analysis which aims at determining the potential
of the EIC to measure gluon shadowing and
anti-shadowing and its sensitivity to
saturation dynamics. 

\begin{figure}[t]
\includegraphics[width=0.52\textwidth]{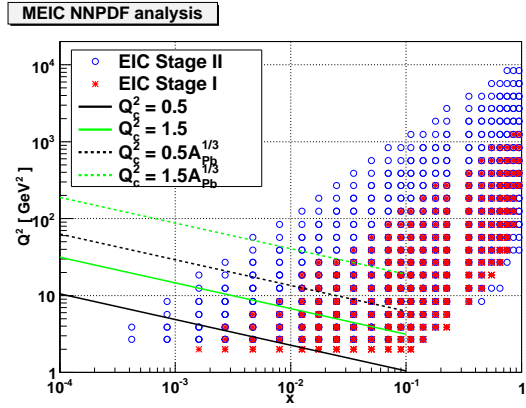}
\caption{\small \label{fig:kin-meic} Kinematical coverage
of the pseudo-data included in the NNPDF analysis of the
EIC Pb cross sections, both for stage I and for stage II. 
Kinematical cuts relevant
to study the onset of non--linear phenomena are also shown.
}
\end{figure}

The input for this analysis is the EIC pseudo data for the inclusive DIS
cross section in two scenarios,
a medium  energy EIC ($\sqrt{s}=12,17,24,32,44$ GeV, denoted by
stage I) and a full energy EIC ($\sqrt{s}=63, 88, 124$ GeV, stage
II). The kinematic coverage is summarized in Fig.~\ref{fig:kin-meic}. 
The pseudo-data was generated starting from $e+p$ and $e+n$ cross
sections computed using the  NNPDF2.0 set~\cite{Ball:2010de}. 
An integrated 
luminosity of 4 fb$^{-1}$ was assumed for all energies, and the
pseudo-data has been
corrected for the expected statistical fluctuations. For most of
the $x$ range the resulting statistical errors are negligible compared
to the assumed 2\% systematic error.
Nuclear effects have been included in 
the
approximation
where
the longitudinal and transverse cross sections 
in Lead ($^{208}$Pb) can be
expressed in terms of the proton cross sections as
\begin{equation}
  \sigma_{T,L}^{\rm Pb}\left( x,Q^2,y \right) =K^{\lambda}_{T,L}\left( x,Q^2,y \right)
  \sigma_{T,L}^{\rm p}\left( x,Q^2,y\right) \ ,
\label{eq:sigma_TL}
\end{equation}
where 
the factors $K$ describe nuclear effects;
the label $\lambda$ sets the intensity of the assumed saturation
effects, and $\lambda=1$ corresponds to the nominal saturation in 
the IP Non-sat model~\cite{Kowalski:2003hm}, 
{\it i.e.},
we assume no saturation for the interaction with the nucleons.
In particular, the $K$-factors in Eq.~(\ref{eq:sigma_TL}) are given by the following piece-wise expression. For small $x \leq 0.01$, $K^{\lambda}_{T,L}$
is given in terms of the dipole cross section  of the IP
Non-sat model.
In the $0.01 \leq x \leq 0.1$ interval, we assume that $K^{\lambda}_{T,L}$
increases linearly from the value given
by the IP Non-sat model at $x=0.01$ up to $K^{\lambda}_{T,L}=1$ at $x=0.1$.
Finally, for $x >0.1$, we assumed that $K^{\lambda}_{T,L}$ is equal to the ratio of the nuclear to
free nucleon structure functions, $F_{2A}(x,Q^2)/[AF_{2N}(x,Q^2)]$ taken
from Ref.~\cite{Eskola:1998df}.
This simple model is intended for our initial studies summarized in
this contribution; a more elaborate model will be considered in the future.

\begin{figure}[t]
\includegraphics[width=0.8\textwidth]{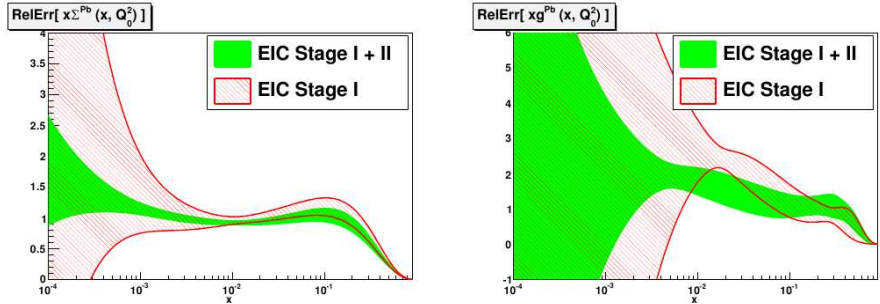}
\caption{
\small \label{fig:pb-pdfs} The quark
singlet (left plot) and the gluon PDFs in 
$^{208}$Pb
(right plot)
at the initial evolution scale $Q_0^2=2$ GeV$^2$, for stage
I and stage I+II.
}
\end{figure}

\begin{figure}[t]
\includegraphics[width=0.40\textwidth]{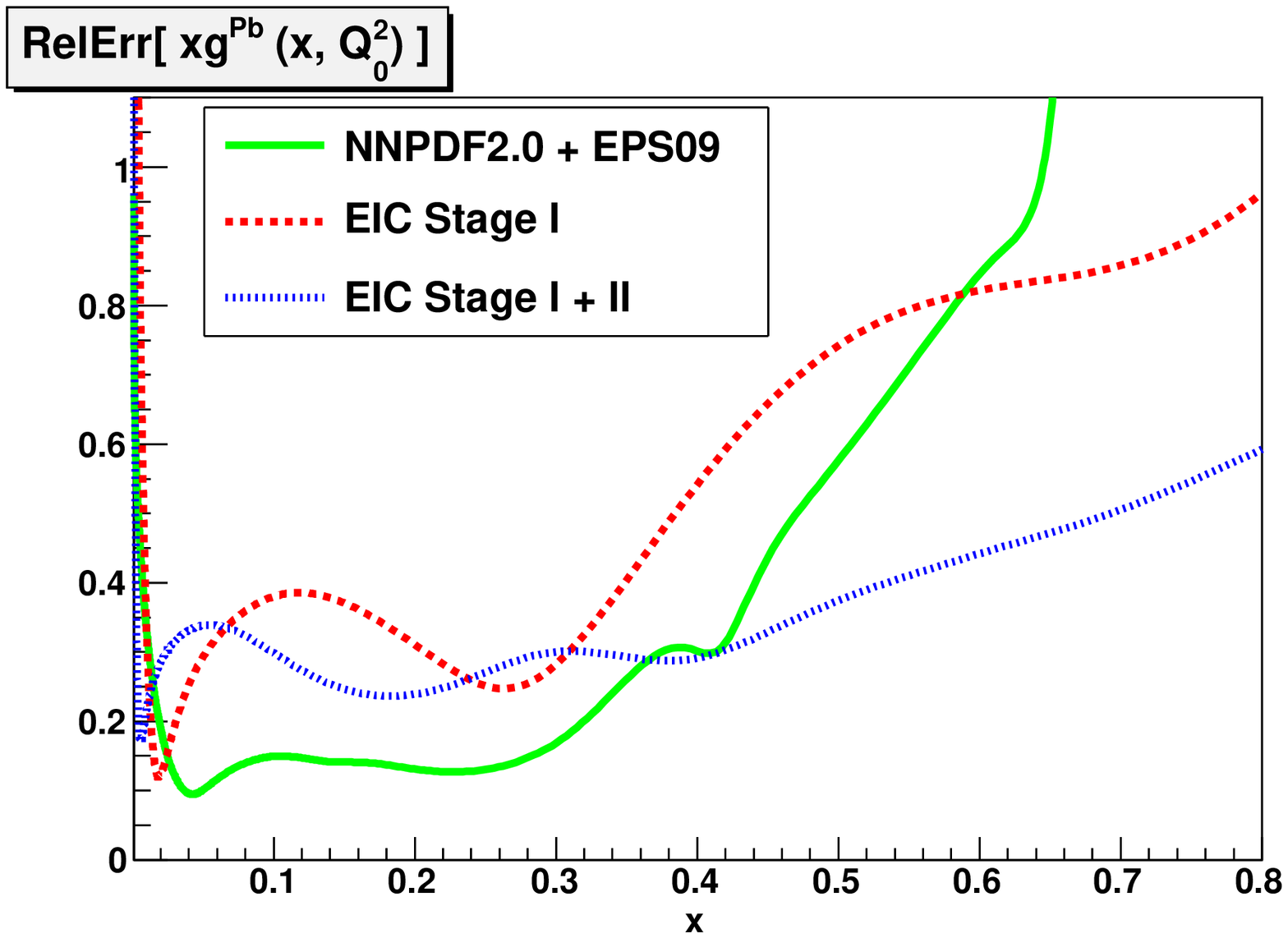}
\includegraphics[width=0.40\textwidth]{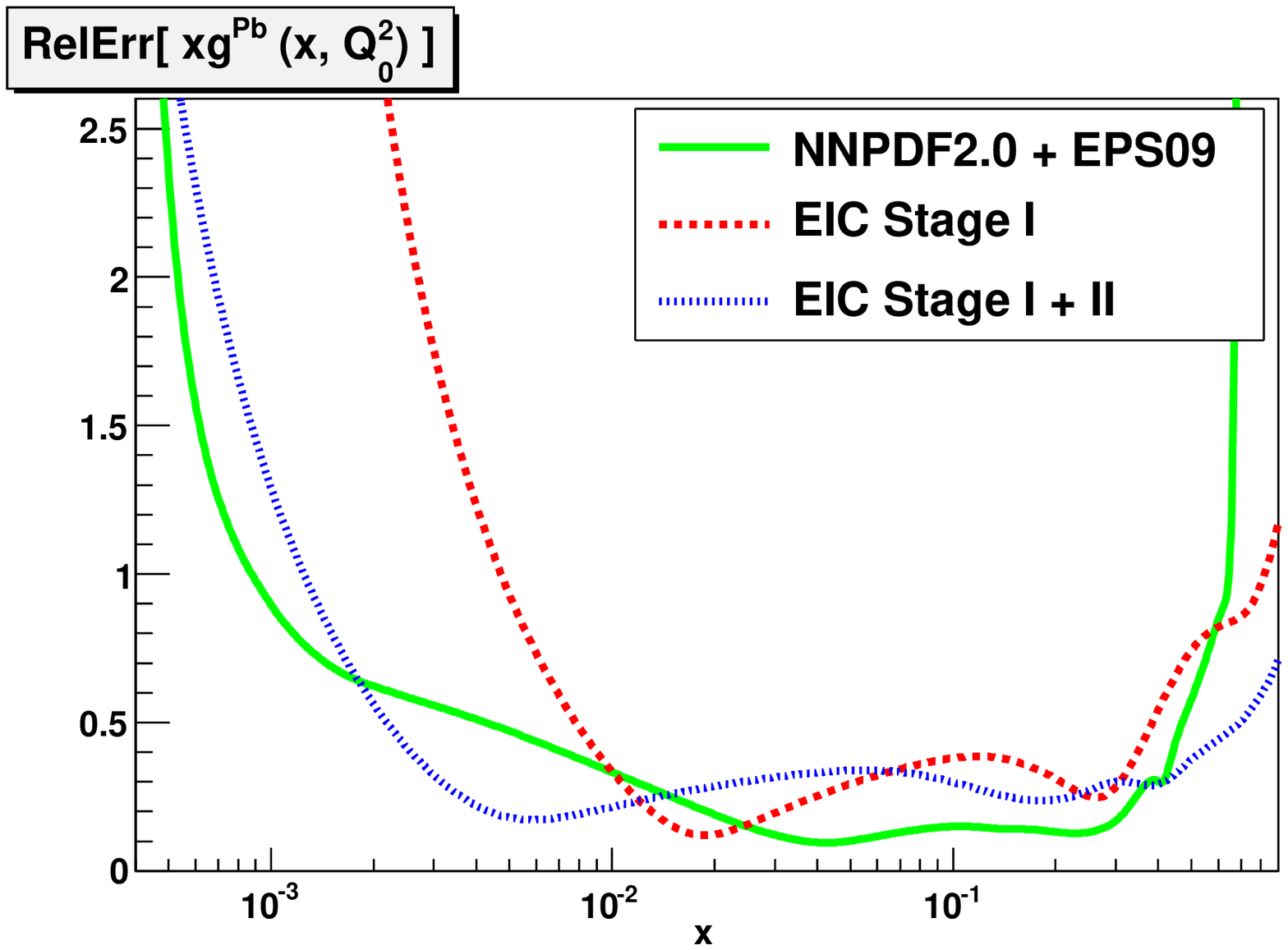}
\caption{\small \label{fig:pb-pdfs-rel} The relative uncertainty in the
 gluon PDF in 
$^{208}$Pb 
at the initial evolution scale $Q_0^2=2$ GeV$^2$, with stage I and
stage I+II data. 
The analogous results for the 
PDFs
in $^{208}$Pb
 using NNPDF2.0+EPS09
parametrizations are also shown.}
\end{figure}

Nuclear parton distributions are then determined by a Next-to-Leading
Order QCD fit of the pseudo-data within the NNPDF
framework~\cite{Ball:2010de,Ball:2008by,Ball:2009mk,Ball:2011mu},
assuming collinear factorization for nuclear targets, and only using
pseudo-data for $^{208}$Pb.
The kinematic cuts used to ensure the validity of DGLAP evolution are
$Q^2\ge 2$ GeV$^2$ and $W^2\ge$ 12.5 GeV$^2$.
In Fig.~\ref{fig:pb-pdfs}, we show  
the singlet and the gluon Lead PDFs at the initial scale $Q^2=2$ GeV$^2$ 
obtained using only stage I data, and then adding the stage II data.
To illustrate the accuracy that the EIC can reach in the determination
of nuclear gluon PDF we show in Fig.~\ref{fig:pb-pdfs-rel} 
their relative uncertainties alongside those of the proton's
NNPDF2.0~\cite{Ball:2010de} combined with those of 
the EPS09 nuclear modifications~\cite{Eskola:2009uj} 
for $^{208}{\rm Pb}$, The NNPDF2.0 and EPS09 
relative uncertainties have been added linearly for a
conservative estimate of the total uncertainty.

The measurement of the nuclear modifications of the gluon
are one of the most important measurements at the EIC, since
this quantity is essentially unknown from present data.
From  Fig.~\ref{fig:pb-pdfs} we see that one can determine
with a reasonable accuracy the gluon shadowing down to $x\sim 10^{-3}$
in stage II and  down to $x\sim 10^{-2}$ in stage I.
The better capabilities of stage II stem both from its
greater lever arm in $Q^2$ and its coverage of smaller
values of $x$, see Fig.~\ref{fig:kin-meic}. In particular,
the precision of 
the determination of the gluon distribution in $^{208}$Pb
in Stage II at small $x$
is comparable to estimates from
global proton fits.
On top of this, at the EIC it will be possible to study gluon
anti-shadowing, 
and 
EMC and Fermi motion effects 
in the gluon channel
with much better 
accuracy than afforded by current global nuclear fits.
We can also see
that EIC will measure accurately the sea quark shadowing, and
that nuclear modifications of light quarks at large $x$ could be
measured a precision similar or even better than for the proton case.  

The 
presented
analysis was based on the validity of collinear
factorization for nuclei, and the validity of linear DGLAP evolution in
$Q^2$. However, at small enough $x$ and $Q^2$, deviations
from linear 
fixed-order
 DGLAP evolution are
expected to appear, e.g., due to small-$x$ resummation 
effects~\cite{Altarelli:2008aj} or
gluon saturation~\cite{Gelis:2008zzc}. 
In Refs.~\cite{Caola:2009iy,Caola:2010cy} a general 
strategy was presented to quantify potential deviations from NLO DGLAP
evolution, which was then applied to proton
HERA data.
In particular,
in
a global PDF fit, deviations from DGLAP in the data can be hidden
in a distortion of parton distributions; however, these can be singled
out by determining undistorted PDF from data in regions where
such effects are 
expected to be
small, evolving them down in the $Q^2$  region 
where deviations are expected to arise and comparing calculations to data
 not used in the PDF determination. 

This approach can be applied as well to the nuclear case. 
From simple theoretical arguments about the energy and atmic number
$A$ dependence of the saturation scale~\cite{Gelis:2008zzc}, 
we expect deviations from linear evolution to appear when
$
  Q^2 \lesssim \bar Q^2 \left( A  \bar{x}/x \right)^{\frac13} \ ,
$
where $\bar x$ is a reference value
(we use $\bar x=10^{-3}$ in our analysis) and
$\bar Q^2$ is the scale where DGLAP evolution at $\bar x$ would be
broken in the proton. 
While saturation models may give an 
indirect
indication of the value of $\bar
Q^2$, we wish to determine this scale in a model independent way
as the scale at which deviations from DGLAP evolution can be
detected from EIC nuclear target 
(pseudo-)data.
 The kinematical
cut above
can also be written as $Q^2 \lesssim Q^2_c x^{-\frac13}$ with 
$Q^2_c$ some constant setting the strength of the deviations from
DGLAP. In Refs.~\cite{Caola:2009iy,Caola:2010cy} the range $Q_c^2\in\lc
0.5,1.5\rc$ GeV$^2$ was considered for the proton case; in the nuclear
case 
one expects that this range should be rescaled by a factor $A_{\rm
  Pb}^{1/3}\approx 6$. 
(Note that nuclear shadowing may reduce this impulse approximation estimate.)
Typical values of these kinematical cuts for
the 
nucleus 
of 
$^{208}$Pb
are shown in  Fig.~\ref{fig:kin-meic}. 

\begin{figure}[t]
\includegraphics[width=0.40\textwidth]{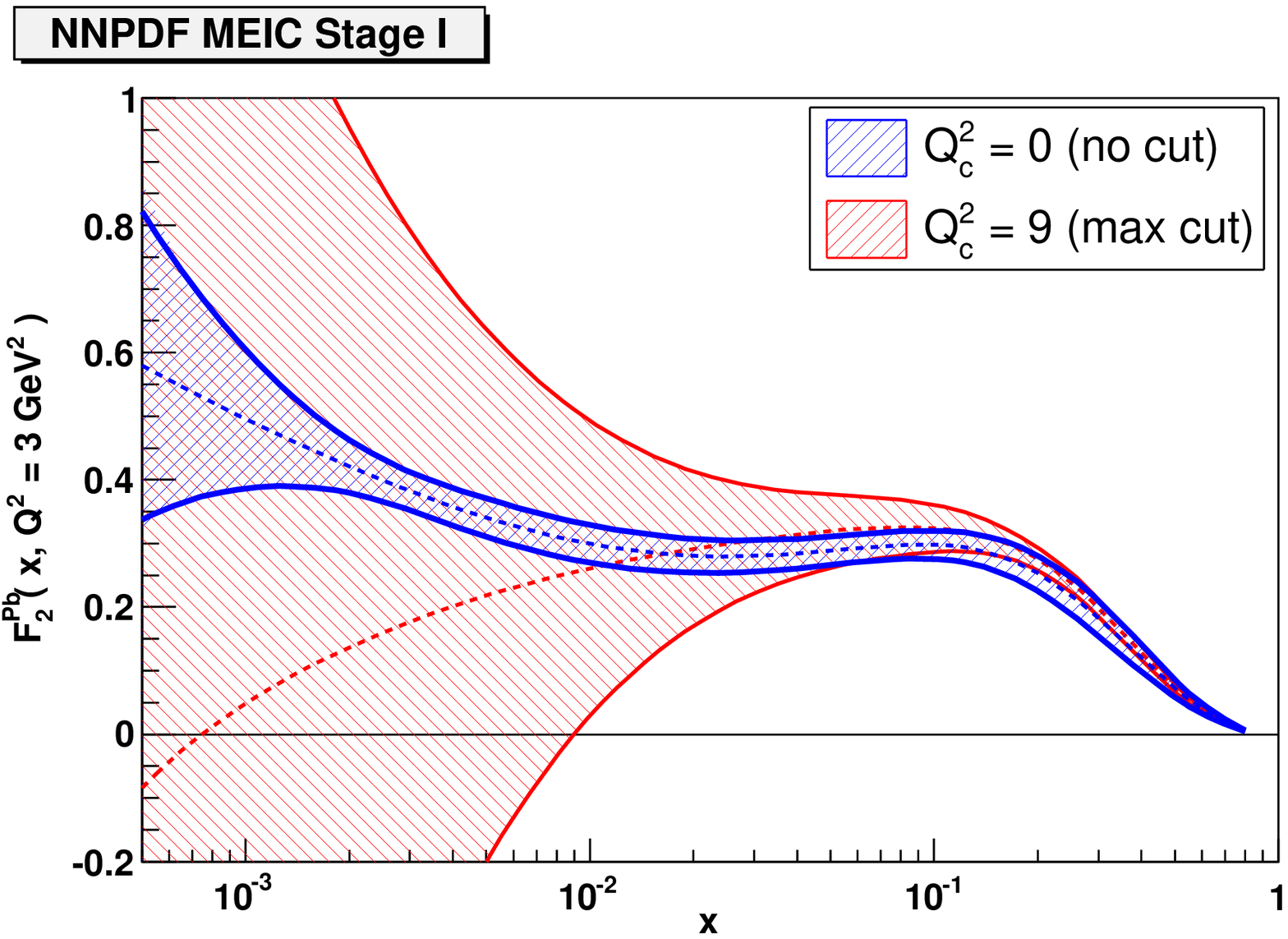}
\includegraphics[width=0.40\textwidth]{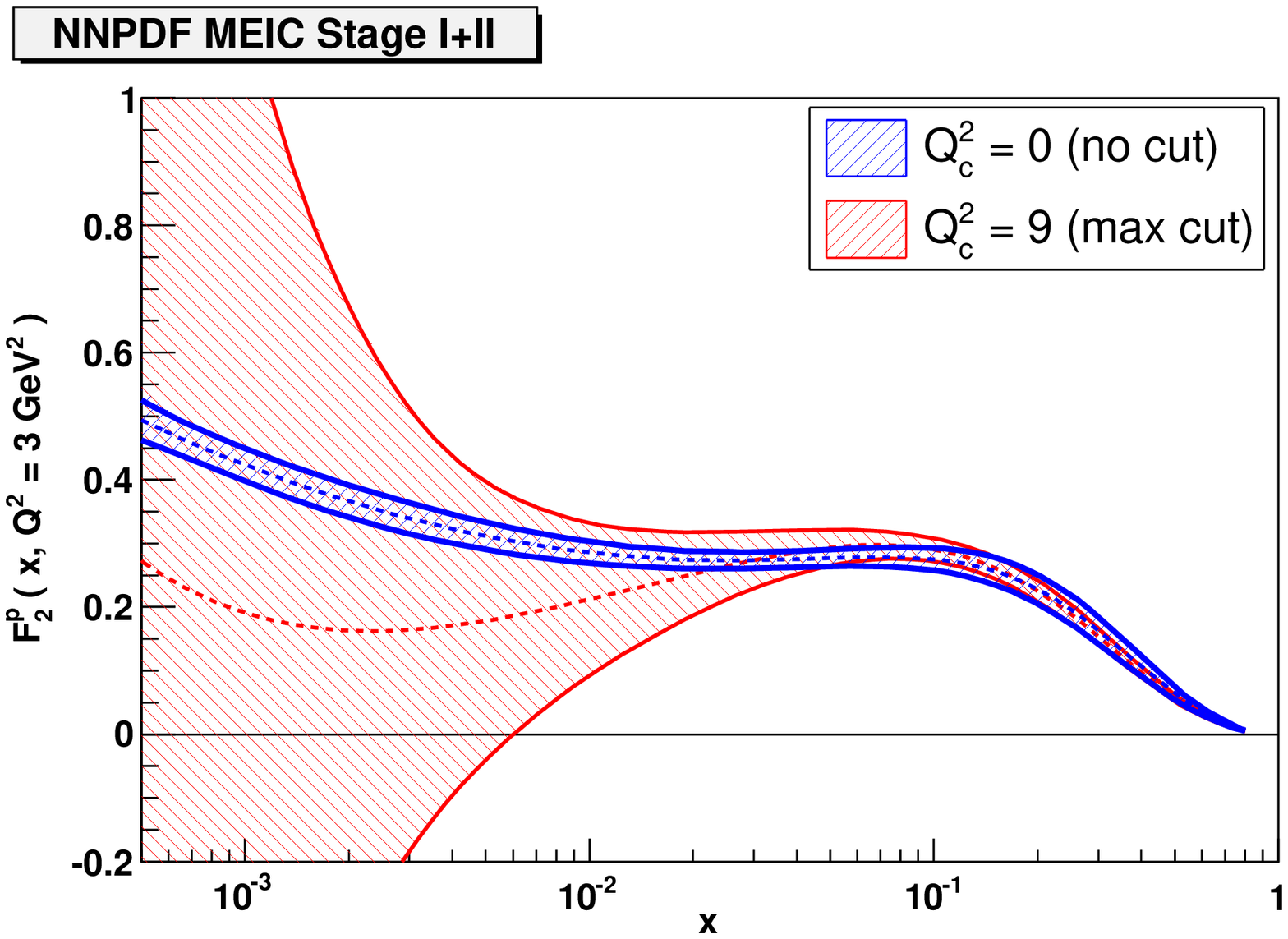}
\caption{\small \label{fig:f2p-cut} 
  The Lead structure function $F_2^{\rm Pb}(x,Q^2)$ at $Q^2=3$ GeV$^2$
  from the analysis of the EIC stage I (left plot) and
  stage I+II (right plot) simulated data with $\lambda=1$, without
  kinematical cuts and with cuts using $Q_c^2=1.5A_{\rm Pb}^{1/3} \sim 9$.
}
\end{figure}

We show in Fig.~\ref{fig:f2p-cut} a representative result
of the fits to the EIC pseudo-data after applying the cut
with $\bar Q^2 =1.5A_{\rm Pb}^{1/3}\sim 9$,
compared to the reference uncut fits to stages I and I+II pseudo-data
with $\lambda=1$.
As expected when data is removed the uncertainties in the
physical observables become much larger, but one can still
see a systematic downwards shift in the central value, which is the
signature of the departure from linear 
evolution~\cite{Caola:2009iy,Caola:2010cy}. Note that this signal is
already apparent with stage I data only, although its statistical
significance might be marginal. Following this preliminary study,
in the future we will present
more detailed and quantitative studies of deviations from DGLAP
in eA collisions at the EIC.


\begin{theacknowledgments}

This work has been supported by the DOE contract DE-AC05-06OR23177,
under which Jefferson Science Associates, LLC operates Jefferson Lab,
and NSF awards No.~0653508 and No.~1002644.

\end{theacknowledgments}



\bibliographystyle{aipproc}   

\begin{thebibliography}{12}
\expandafter\ifx\csname natexlab\endcsname\relax\def\natexlab#1{#1}\fi
\providecommand{\enquote}[1]{``#1''}
\expandafter\ifx\csname url\endcsname\relax
  \def\url#1{\texttt{#1}}\fi
\expandafter\ifx\csname urlprefix\endcsname\relax\def\urlprefix{URL }\fi
\providecommand{\eprint}[2][]{\url{#2}}

\bibitem[Ball et~al.(2010)]{Ball:2010de}
R.~D. Ball, et~al., \emph{Nucl. Phys.} \textbf{B838}, 136--206 (2010),
  \eprint{arXiv:1002.4407}.

\bibitem[Kowalski and Teaney(2003)]{Kowalski:2003hm}
H.~Kowalski, and D.~Teaney, \emph{Phys. Rev.} \textbf{D68}, 114005 (2003),
  \eprint{hep-ph/0304189}.

\bibitem[Eskola et~al.(1999)]{Eskola:1998df}
K.~J. Eskola, V.~J. Kolhinen, and C.~A. Salgado, \emph{Eur. Phys. J.}
  \textbf{C9}, 61--68 (1999), \eprint{hep-ph/9807297}.

\bibitem[Ball et~al.(2009{\natexlab{a}})]{Ball:2008by}
R.~D. Ball, et~al., \emph{Nucl. Phys.} \textbf{B809}, 1--63
  (2009{\natexlab{a}}), \eprint{arXiv:0808.1231}.

\bibitem[Ball et~al.(2009{\natexlab{b}})]{Ball:2009mk}
R.~D. Ball, et~al., \emph{Nucl. Phys.} \textbf{B823}, 195--233
  (2009{\natexlab{b}}), \eprint{arXiv:0906.1958}.

\bibitem[Ball et~al.(2011)]{Ball:2011mu}
R.~D. Ball, et~al., \emph{Nucl. Phys.} \textbf{B849}, 296--363 (2011),
  \eprint{arXiv:1101.1300}.

\bibitem[Forte(2010)]{Forte:2010dt}
S.~Forte, \emph{Acta Phys. Polon.} \textbf{B41}, 2859--2920 (2010),
  \eprint{arXiv:1011.5247}.

\bibitem[Eskola et~al.(2009)]{Eskola:2009uj}
K.~J. Eskola, H.~Paukkunen, and C.~A. Salgado, \emph{JHEP} \textbf{04}, 065
  (2009), \eprint{arXiv:0902.4154}.

\bibitem[Altarelli et~al.(2008)]{Altarelli:2008aj}
G.~Altarelli, R.~D. Ball, and S.~Forte, \emph{Nucl. Phys.} \textbf{B799},
  199--240 (2008), \eprint{arXiv:0802.0032}.

\bibitem[Gelis(2008)]{Gelis:2008zzc}
F.~Gelis, \emph{Acta Phys. Polon.} \textbf{B39}, 2419--2454 (2008).

\bibitem[Caola et~al.(2010)]{Caola:2009iy}
F.~Caola, S.~Forte, and J.~Rojo, \emph{Phys. Lett.} \textbf{B686}, 127--135
  (2010), \eprint{arXiv:0910.3143}.

\bibitem[Caola et~al.(2011)]{Caola:2010cy}
F.~Caola, S.~Forte, and J.~Rojo, \emph{Nucl. Phys.} \textbf{A854}, 32--44
  (2011), \eprint{arXiv:1007.5405}.

\end{thebibliography}


\end{document}